\documentclass[sigconf]{acmart}

\AtBeginDocument{%
  \providecommand\BibTeX{{%
    \normalfont B\kern-0.5em{\scshape i\kern-0.25em b}\kern-0.8em\TeX}}}

\usepackage[T1]{fontenc}
\usepackage{graphicx}

\usepackage{booktabs} 
\usepackage{color}
\usepackage{enumitem}

\usepackage{multirow}
\usepackage{tabularx}
\usepackage{xspace}

\usepackage{tcolorbox}

\usepackage{pifont}

\newcommand{\boldheading}[1]{%
  \par\addvspace{0.5\baselineskip}%
  \noindent\textbf{\emph{#1.}}%
}

\definecolor{varblue}{RGB}{0, 102, 204}
\definecolor{vargreen}{RGB}{0, 153, 0}

\newcommand\RecQuest{\textsc{RecQuest}\xspace}

\newtcolorbox{promptbox}{%
    width=0.98\linewidth,
    colback=black!5!white, 
    colframe=black!75!white, 
    arc=1mm, 
    boxrule=0.8pt, 
    top=0.5mm, bottom=1mm, left=2mm, right=2mm, 
    boxsep=1mm,
    sharp corners=south
}

\copyrightyear{2026}
\acmYear{2026}
\setcopyright{cc}
\setcctype{by}
\acmConference[ICTIR '26]{Proceedings of the 2026 International ACM SIGIR Conference on Innovative Concepts and Theories in Information Retrieval (ICTIR)}{July 25, 2026}{Melbourne, VIC, Australia}
\acmBooktitle{Proceedings of the 2026 International ACM SIGIR Conference on Innovative Concepts and Theories in Information Retrieval (ICTIR) (ICTIR '26), July 25, 2026, Melbourne, VIC, Australia}
\acmDOI{10.1145/3805713.3820438}
\acmISBN{979-8-4007-2600-2/2026/07}

\begin{document}
\title{RecQuest: Towards Estimating User Domain Knowledge in Conversational Recommender Systems}

\author{Ivica Kostric}
\affiliation{%
  \institution{University of Stavanger}
  \city{Stavanger}
  \country{Norway}
}
\email{ivica.kostric@uis.no}

\author{Ujwal Gadiraju}
\affiliation{%
  \institution{Delft University of Technology}
  \city{Delft}
  \country{Netherland}
}
\email{u.k.gadiraju@tudelft.nl}

\author{Krisztian Balog}
\affiliation{%
  \institution{University of Stavanger}
  \city{Stavanger}
  \country{Norway}
}
\email{krisztian.balog@uis.no}

\renewcommand{\shortauthors}{Ivica Kostric, Ujwal Gadiraju, and Krisztian Balog}
\begin{abstract}
The ideal conversational recommender system (CRS) acts like a savvy salesperson, adapting its language and suggestions to a user's expertise level. However, most current systems treat all users as experts, leading to frustrating and inefficient interactions when users are unfamiliar with a domain. Systems that can adapt their conversational strategies to a user's knowledge level stand to offer a much more natural and effective experience. To enable such adaptation, a CRS must first be able to estimate a user’s domain knowledge from interaction signals. Yet, accurately estimating knowledge typically requires tailored interactions to elicit those signals in the first place, creating a fundamental chicken-and-egg problem. In this work, we take a first step toward breaking this dependency by introducing a new task: estimating user domain knowledge directly from conversational transcripts. A key obstacle to such estimation is the lack of suitable data; to our knowledge, no existing dataset captures the conversational behaviors of users with varying levels of domain knowledge. Furthermore, in most dialogue collection protocols, users are free to express their own preferences, which tends to concentrate on popular items and well-known features, offering little insight into how novices explore or learn about unfamiliar features. To address this, we design RecQuest, a game-with-a-purpose data collection protocol that elicits varied expressions of knowledge while using a target-aware CRS to guide interactions, release the resulting dataset, and provide baseline methods and analyses to support future work on user-knowledge-aware CRS.
\end{abstract}



\maketitle
\section{Introduction}
\label{introduction}

\begin{figure}[t]
  \centering
  \includegraphics[width=.95\linewidth]{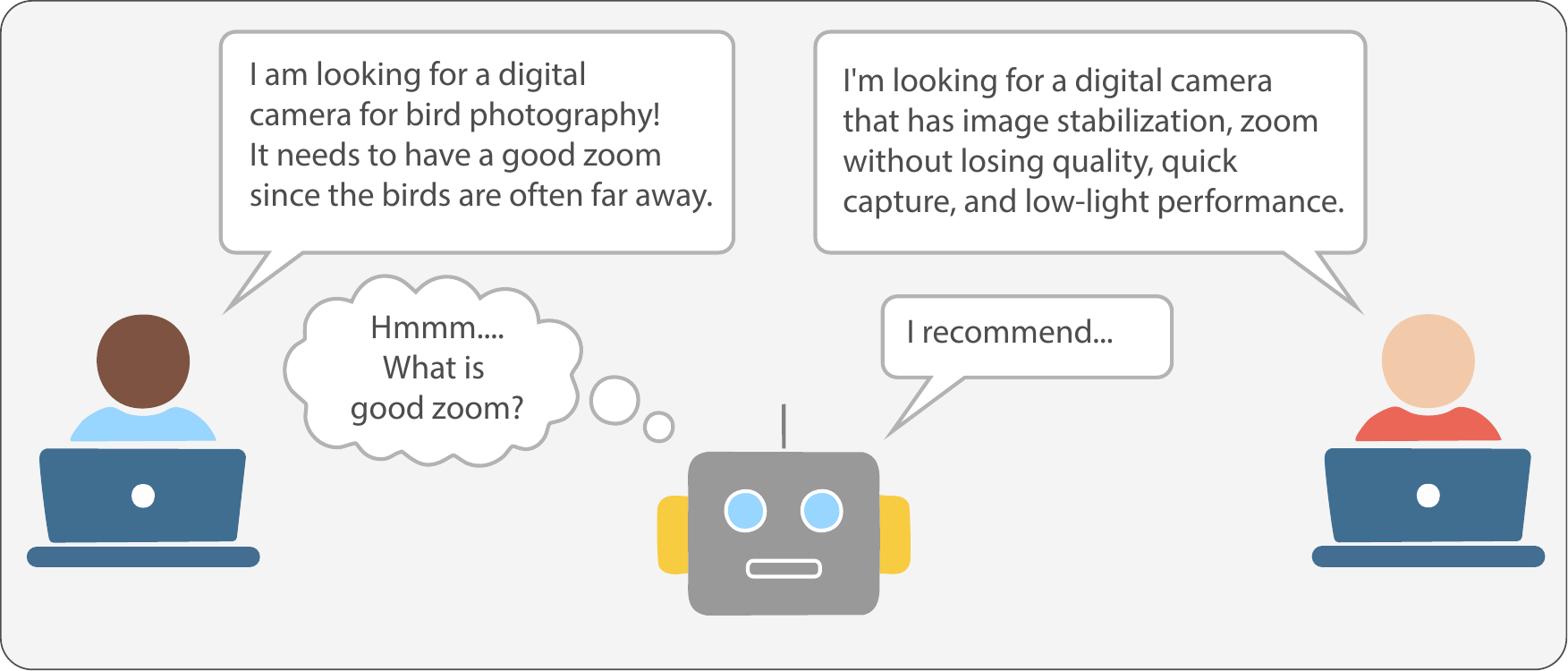}
  \caption{An example from our collected dataset illustrating the different levels of product specification between a novice and an expert while describing their needs for a digital camera purchase. Novices tend to describe goals and contexts, whereas experts tend to specify technical constraints and components more clearly.}
  \label{fig:teaser}
\end{figure}

Conversational recommender systems (CRS) address the challenge of helping users find relevant items in vast catalogs by engaging them in interactive dialogues, typically involving preference elicitation based on item attributes~\citep{Jannach:2022:ACM}. Their effectiveness relies heavily on accurately capturing user preferences and interpreting their expressions about desired attributes~\citep{Pramod:2022:Expert}.
However, most current CRS approaches typically assume that users understand item attributes clearly and can directly map them to their preferences~\citep{Zhao:2022:WWW, Lin:2023:WWW, Zhang:2025:ACM}. In contrast, \citet{Kostric:2024:TORS} propose a usage-oriented preference elicitation strategy, specifically targeting users with low-domain knowledge. In isolation, both approaches can lead to limited and inefficient interactions, as users differ significantly in domain knowledge, vocabulary, and preferred interaction style. As illustrated in Figure \ref{fig:teaser}, a novice might express their needs differently than an expert when buying a digital camera. 
As a result, systems that rely on a single interaction strategy force users to adapt to the system's language. However, recent work shows that task performance and user satisfaction improve significantly when a CRS instead adapts its interaction style to the user's domain knowledge~\citep{Kostric:2025:UMAP}.

To bridge this gap and enable the creation of more adaptive systems, we introduce the novel task of estimating user domain knowledge from conversations, allowing a CRS to tailor its interaction strategy---from preference elicitation to the presentation of recommendations and explanations. While empirical studies in information retrieval and recommender systems have shown that domain expertise shapes user interactions~\citep{White:2009:WSDM, Mao:2018:TOIS}, estimating this knowledge from multi-turn dialogues remains a novel problem in the CRS setting.

A major obstacle in advancing the task of user domain knowledge estimation is the lack of datasets that capture such domain knowledge within a conversational context. Existing CRS datasets rely on open-ended protocols in which participants discuss only the attributes they are familiar with~\citep{Joko:2024:SIGIR, Radlinski:2019:SIGDIAL, Li:2018:NIPS}. This approach leads to conversations focused primarily on well-known aspects of the domain, providing limited evidence about how users express needs, ask questions, or refine preferences when encountering unfamiliar aspects.
Given these challenges, a critical question is: \emph{(\textbf{RQ1}) How can we design a data collection protocol that elicits signals of domain knowledge through engaging and natural conversations?} We address this by developing a game-with-a-purpose (GWAP)\footnote{Here, ``game'' means an incentivized task with goals and rules, designed to elicit useful data from participants, rather than a game-theoretic model of strategic interaction.}~\citep{VonAhn:2006:Computer,balayn2022ready} protocol that combines open-ended interaction with a target-oriented objective. Specifically, participants are assigned a hidden target item, presented only as a concise background narrative. During the interaction, they must interact with the CRS to identify the item that best matches the narrative by providing preference statements, responding to elicitation prompts, and discussing the system's recommendations. This design is distinct from both known-item search~\citep{Lee:2006:ASIST} and open-ended preference elicitation, while combining elements of each: participants are looking for a particular item without knowing its exact identity, while the open-ended protocol encourages the articulation, refinement, and clarification of preferences that expose signals of domain knowledge.

An inherent challenge in estimating user domain knowledge is determining the ground truth. This leads us to ask: \emph{(\textbf{RQ2}) How do objective knowledge scores compare to user self-assessments, and which provides a more reliable ground truth?} We address this by contrasting participants' self-assessed confidence ratings with their performance on a domain-specific technical questionnaire. We then determine the more robust ground truth by analyzing which measure better predicts actual task success, thereby identifying the calibration biases that undermine self-reported expertise.

With the ground truth established, we next examine which observable signals distinguish users with varying domain knowledge. Specifically, we ask: \emph{(\textbf{RQ3}) What identifiable traces differentiate expert behavior from novice behavior, and to what extent do these patterns make users with different knowledge levels distinguishable in practice?} To answer this, we analyze the collected dialogues for measurable linguistic and interactional markers, including utterance length, vocabulary, and the frequency and specificity of the phrases used. Based on prior work~\citep{White:2009:WSDM, Mao:2018:TOIS}, we expect that knowledgeable users articulate their needs through more precise, attribute-oriented language, whereas users with limited domain knowledge rely on high-level or usage-based descriptions and exhibit more frequent clarification behavior.

Finally, we investigate the feasibility of automatic knowledge estimation from dialogue. Specifically, \emph{(\textbf{RQ4}) How effectively can large language models (LLMs) estimate users' domain knowledge from conversational data?} To answer this, we examine whether these models can infer knowledge levels directly from raw dialogue transcripts. This involves assessing how well-separated the knowledge categories are, and whether users at different expertise levels exhibit sufficiently distinct linguistic, behavioral, or interactional markers to enable robust classification. 

In summary, our contributions are as follows:
\begin{itemize}[nosep, leftmargin=*]
    \item We introduce the task of estimating user domain knowledge in conversational recommenders directly from dialogues.
    \item We design \RecQuest, a game-with-a-purpose data collection protocol that uses background narratives and attribute clues to elicit natural conversations rich in domain knowledge signals.
    \item  We collect and release a novel CRS dataset capturing these domain-specific interactions. We note that this dataset is intended primarily to validate the proposed collection protocol and support initial feasibility analyses, rather than to serve as a large-scale supervised training benchmark.
    \item We conduct a detailed analysis of the collected dialogues to uncover the behavioral and linguistic patterns that distinguish users with different knowledge levels.
    \item We establish LLM-based baselines for estimating user domain knowledge from conversational data, and analyze the separability of knowledge levels on this challenging task.
\end{itemize}
The developed CRS, study protocols, and collected dialogues are made available at: \url{https://github.com/iai-group/crs-knowledge}.

\section{Related Work}
\label{rel_work}

User domain knowledge has been previously investigated in the context of web search \citep{White:2009:WSDM, Tabatabai:2005:LISR, Mao:2018:TOIS}.  Prior work differs mainly in how expertise is determined and modeled.  The two most common approaches for determining user knowledge are subjective self-assessments and objective knowledge tests.  Self-assessment studies ask users to report their familiarity or perceived expertise, most often distinguishing between two and five categories~\citep{Noh:2023:CHI, Ferrod:2021:UMAP, McAuley:2013:WWW}. Objective approaches instead---also popularly employed in the `\textit{search as learning}' realm---rely on domain quizzes or knowledge tests whose scores define knowledge levels \citep{Kiseleva:2015:arXiv, Yu:2018:SIGIR, Mao:2018:TOIS, Zhang:2011:SIGIR, Zhang:2015:JASIST, collins2016assessing,gadiraju2018analyzing}. Research based on both approaches consistently finds that experts issue more structured and technical queries, use more precise and attribute-oriented vocabulary, reformulate less, while novices rely more on general, case-oriented descriptions and ask more clarification questions \citep{Tabatabai:2005:LISR, Noh:2023:CHI}. Prior work on adaptive preference elicitation further shows that user characteristics and domain familiarity shape interaction behavior and satisfaction, with more knowledgeable users responding better to structured, attribute-focused elicitation, and less knowledgeable users benefiting more from guided and example-based approaches~\citep{Knijnenburg:2009:RecSys}.

Our work is situated at the intersection of two key research areas. We first review the main paradigms of CRS, as they provide the architectural context for the system we developed for our data collection. We then critically examine existing dialogue collection methods, highlighting the gap our protocol directly addresses.

\boldheading{Conversational Recommender Systems}
Research on CRS is broadly categorized into two main groups.  \emph{Attribute-based CRS} cast the dialogue as a slot-filling exercise~\citep{Lei:2020:WSDMa, Christakopoulou:2016:KDD, Bernard:2024:WSDM}, eliciting preferences from users about items or item attributes with fixed templates. 
\emph{Generation-based CRS}, in contrast, generates free-form responses while recommending items, generally utilizing an end-to-end architecture that integrates the conversation and the recommendation components~\citep{Zhou:2020:COLING}.
Our work is relevant to both paradigms, as estimating user knowledge can, among others, help tailor the elicitation strategy and the explanatory language.
Our work employs a hybrid CRS that combines attribute-based preference elicitation with natural language generation for recommendations and explanations. The system maintains an internal representation of user preferences and supports interactions entirely in natural language.

\boldheading{Dialogue Collection Methods}
Significant effort has been dedicated to collecting datasets that capture how users interact with conversational recommenders~\citep{Li:2018:NIPS, Hayati:2020:EMNLP, Shah:2018:NAACL, Radlinski:2019:SIGDIAL, Joko:2024:SIGIR, Bernard:2023:SIGIR, Budzianowski:2018:EMNLP, Rastogi:2020:AAAI, Kang:2019:EMNLP}. Influential examples like  ReDial~\citep{Li:2018:NIPS} and INSPIRED~\citep{Hayati:2020:EMNLP} are created as a conversation between two participants, where one user (the ``seeker'') states open-ended preferences and another (the ``recommender'') suggests items that fit those preferences. However, these open-ended protocols often yield shallow conversations, with seekers frequently accepting recommendations uncritically. More importantly for our work, these protocols do not explicitly control for or capture user domain knowledge, making it difficult to study its effects.
While other datasets use coached human-human protocols to emphasize naturalness (e.g., CCPE-M~\citep{Radlinski:2019:SIGDIAL}, MG-ShopDial~\citep{Bernard:2023:SIGIR}) or focus on human-machine settings (e.g., Schema-Guided Dialogue~\citep{Rastogi:2020:AAAI}, MultiWOZ~\citep{Budzianowski:2018:EMNLP}), none explicitly incorporates user domain knowledge or controls for expertise.

\section{Problem Statement}
\label{sec:problem}

The realization of a truly adaptive conversational recommender system presents a fundamental ``chicken and egg'' problem. Ideally, a CRS should act like a savvy salesperson, adapting its language and suggestions to each user's level of expertise. However, to effectively tailor these interactions, the system must first estimate the user's domain knowledge; yet, accurate estimation often requires tailored interactions---such as specific elicitation strategies---to surface those knowledge signals in the first place. We aim to break this co-dependency by establishing the necessary groundwork for knowledge estimation independent of complex adaptation strategies. In this work, we focus on the first half of this challenge: designing a protocol to capture natural variations in domain knowledge and introducing baseline methods to estimate it. By isolating the estimation task, we provide the resources required to build systems that can eventually close the loop and personalize the experience dynamically.

Following prior work on modeling knowledge, we adopt a discretized representation, where users are grouped into a small number of categories~\citep{Brusilovsky:2007:Springer}. Specifically, we categorize users into three relative levels of knowledge: \texttt{low}, \texttt{medium}, and \texttt{high}. Low domain knowledge is characterized by infrequent and superficial interactions with the domain, where users have limited exposure and may lack a precise technical vocabulary or jargon. In contrast, high domain knowledge corresponds to frequent, hobby-like engagement, where users are more familiar with detailed attributes and exhibit a refined understanding. Medium knowledge falls between these extremes, indicating occasional but shallow interactions.

\section{The RecQuest Data Collection Protocol}
\label{sec:game}

To study knowledge estimation, we require data that captures the nuanced conversational behaviors of users with varying levels of expertise---a resource absent in existing open-ended datasets.
Current data collection protocols typically elicit unconstrained preferences, which tend to focus on popular, widely known items and allow users to accept recommendations with minimal scrutiny. In such settings, successful interaction does not require users to articulate domain-specific constraints, use lesser-known domain terminology, or explore trade-offs, making domain expertise largely irrelevant and consequently difficult to observe. 
To address this, we introduce \RecQuest, a game-with-a-purpose protocol that incentivizes users to explore domain boundaries by assigning them a specific target item hidden within a background narrative. This section details the game mechanics, the item collection process across five distinct domains (bicycles, laptops, digital cameras, running shoes, and smartwatches), and the CRS architecture designed to support these goal-directed interactions.

\subsection{Game Mechanics}
\label{data:game}

\begin{figure*}[t]
  \centering
  \includegraphics[width=.85\textwidth]{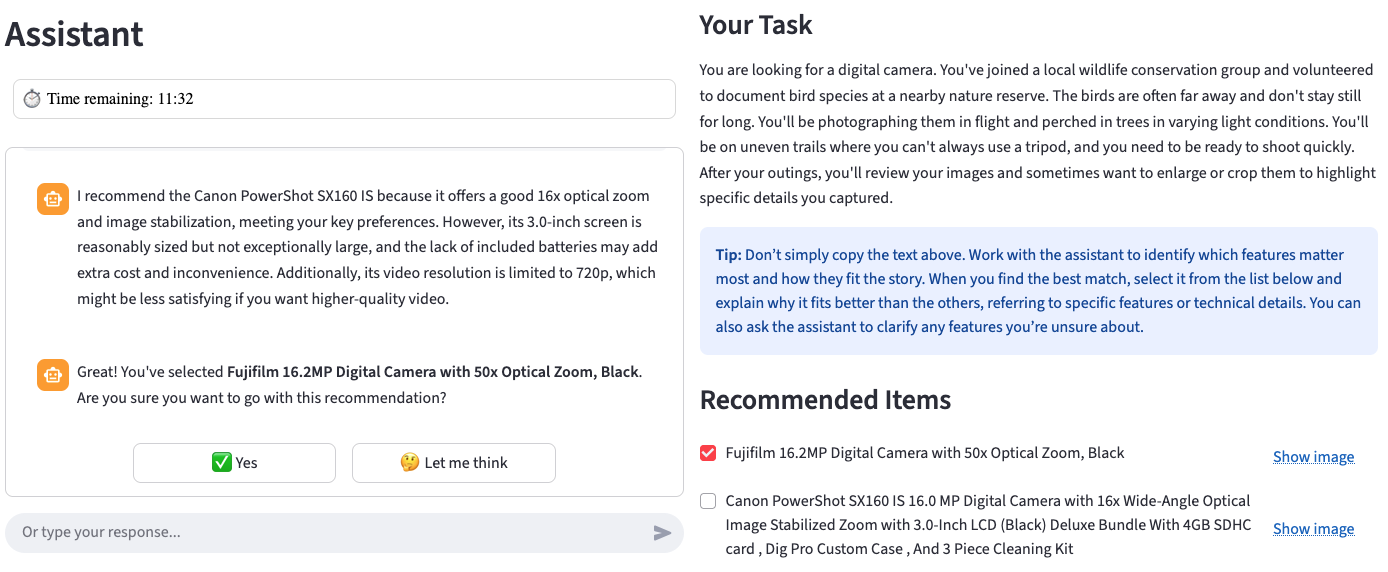}
  \caption{Screenshot of the game in action, showing a user who has obtained several recommendations during the conversation and is ready to make a final selection.}
  \label{fig:screenshot}
\end{figure*}

\boldheading{Game Start} 
At the start, participants see a chat interface and a brief background story setting up a use scenario (Fig.~\ref{fig:screenshot}). The story is generated from a specific item in the curated domain collection to anchor the conversation context without revealing the item itself. Each target item has two background-story variants: a concise (5 feature constraints, 80–100 words) and a more detailed version (10 feature constraints, 150–200 words); see Section~\ref{sec:data_collection:generating_game_states} for details. This enables us to investigate whether the richness of contextual cues affects how participants formulate preferences or describe product features. 
Participants are instructed to find the item that best fits the scenario by engaging in a turn-based dialogue with the CRS.

\boldheading{Participant Turns}
Participant turns are open-ended, allowing for various conversational strategies. 
Participants may state preferences, answer/ask questions, or critique recommendations. 
The ability to ask about an item’s properties allows them to probe which constraints or features are satisfied---much like in real-world interactions between customers and salespeople. Feedback on recommendations is optional. Each recommendation includes an image and a brief system-generated explanation of its fit to the participant’s stated needs. Participants may ask follow-up questions, refine their needs, or ignore the suggestion, using their own vocabulary.

A valid session requires the CRS to have produced at least two recommendations, which remain visible and can be selected at any point. Selecting an item triggers a confirmation prompt, asking whether they are certain of their choice, followed by a request for a brief feature-based justification. Participants are incentivized to identify the correct item by a monetary reward, fostering realistic behavior and increasing the external validity of the collected data.

\boldheading{CRS Turns}
On each system turn, the CRS performs one of three possible actions: asking an elicitation question, answering a participant's question, or providing a recommendation. All actions are generated by a target-aware conversational agent guided by a mix of rule-based heuristics and LLM output.
\begin{itemize}[nosep, leftmargin=*]
    \item \emph{Elicitation.}
When a participant expresses several preferences at once, the CRS restricts them to at most three constraints. If more are listed, the system asks which ones are most important before continuing. This keeps early turns focused and prevents participants from giving the entire need description in a single turn.
Elicitation questions are open-ended but specific to attributes present in the domain’s schema (e.g., \emph{``Is weight or range more important for your use?''}). Thus, they maintain engagement through incremental exchanges. 
    \item \emph{Answering questions.}
When the participant asks about item attributes or domain facts, the CRS responds directly. All factual answers are LLM-generated and framed relative to the target item when applicable, so that information indirectly guides the user toward it without revealing the item’s identity.
    \item \emph{Recommendation.}
A recommendation turn presents a single item from the domain’s curated collection accompanied by an image and a short natural-language explanation. The explanation is generated through a two-step LLM procedure: (i) the system identifies key differences between the candidate item and the hidden target, and (ii) produces an explanation that highlights how those differences relate to the stated preferences (example in Fig. \ref{fig:screenshot}). The system may suggest a previously recommended item if new preferences align with it.
\end{itemize}

\boldheading{Game End}
A session ends when the participant confirms a selection. 
The session length is a configurable parameter (limited to 15 minutes in our work). If the time expires, a soft grace period begins, allowing participants to finalize a selection from the recommendations already shown. 
On confirmation, participants provide a brief, feature-based justification, after which the dialogue concludes.

\subsection{Item Collection}
We curate item collections for each of the five domains used in \RecQuest—bicycles, laptops, digital cameras, running shoes, and smartwatches—using data from the Amazon Review Dataset~\citep{hou2024bridging}. Each domain represents a semantically coherent item category with its own attribute structure, user intent patterns, and relevance criteria. This multi-domain setup allows us to evaluate knowledge estimation performance both within and across heterogeneous product types, providing a controlled yet diverse testbed for \RecQuest. For every domain, we extract metadata, structured attributes, and item descriptions to create representations of available items. To ensure meaningful interaction, duplicates and highly similar items are removed using similarity-based filtering over item descriptions, resulting in diverse collections of distinguishable items.

\subsection{CRS Architecture}
\label{data:flow}

\begin{figure*}[t]
  \centering
  \includegraphics[width=0.9\textwidth]{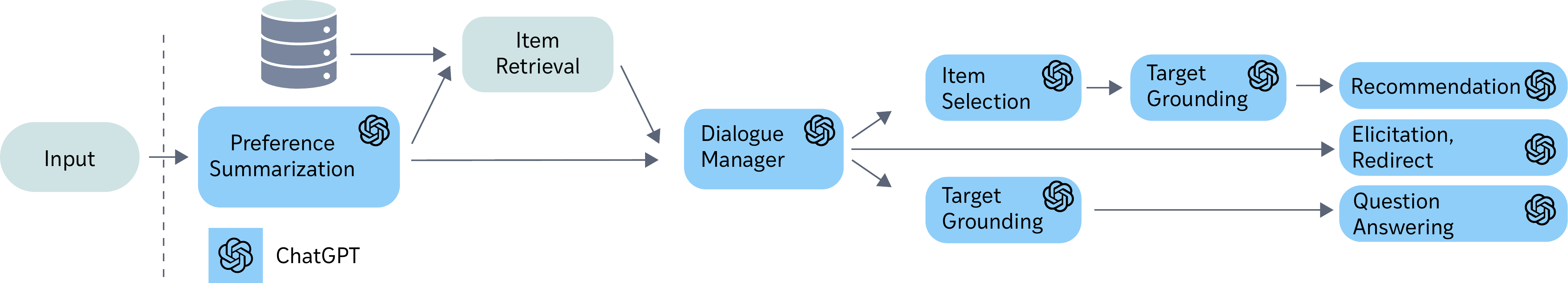}
  \caption{An overview of the CRS architecture, with the flow of data between different components.}
  \label{fig:architecture}
\end{figure*}

Figure~\ref{fig:architecture} shows the CRS architecture and data flow between its core components. At each participant's turn, the system processes the input, updates its internal state, and determines the next intent before producing a response. 

\boldheading{Preference Summarizer} The CRS maintains a structured JSON-like representation of atomic preference statements. After each turn, an LLM-based preference summarizer updates this record by merging newly mentioned constraints with existing ones, ensuring all active needs are captured concisely. 

\boldheading{Item Retrieval} A retrieval step is triggered when the preference record contains at least two statements or when new ones are added. Dense vector representations of items are pre-computed and held in memory. Current preferences are encoded to retrieve relevant candidates. If $\geq 3$ preferences appear in a single turn, the system defers retrieval and asks the participant to prioritize which constraints matter most before continuing. This mechanism is designed to promote multi-turn interaction, encouraging the gradual refinement of needs through dialogue rather than attempting to complete the task in a single turn.

\boldheading{Decision Module} After preference updating and retrieval, the Decision Module applies a hybrid LLM and rule-based policy. It selects the next conversational intent based on the preference list, available recommendations, and the last message. It chooses one of four intents: redirect, answer, recommend, or elicit. \emph{Redirect} is triggered when the dialogue drifts outside the designated domain. \emph{Answer} is selected when the user asks factual or conversational questions. \emph{Recommendation} is selected only if the LLM predicts recommendation and the following rule-based constraints are satisfied: retrieved candidates are available, the preference state contains new information, and at least three preferences have been stated; otherwise, the system falls back to \emph{Elicitation} to gather additional constraints or clarify vague statements.
Once the intent is selected, the system generates a response through specialized LLM modules. 

\boldheading{Response Generation} \emph{Recommendation} turns involve a multi-step process: an LLM selects the best candidate from the top-10 retrieved items given the current preferences and the latest exchange.  
A second model performs \emph{target grounding} by comparing the selected candidate to the hidden target item, identifying which differences are relevant to the participant’s stated needs. This  ensures that explanations focus on meaningful contrasts. A third model generates the final natural language response that presents the recommended item and integrates the contrastive explanation. The \emph{Answering} intent also uses target grounding, particularly for follow-up questions about recommended items. \emph{Redirection} and \emph{Elicitation} use a single generation step to produce a contextually appropriate response. 

\section{Data Collection}
\label{sec:data_collection}

Using the \RecQuest protocol, we conducted a comprehensive study to collect a dataset of dialogues rich in domain knowledge signals. Crucially, our goal was not to construct a large-scale training resource, but rather to validate the protocol and determine whether it effectively elicits observable signals of domain knowledge in conversation. The study followed a four-stage flow designed to balance participant expertise across domains. This section presents a descriptive analysis of the collected data as well as an assessment of the participants' subjective experiences to validate the protocol's engagement and naturalness.

\begin{table*}[t]
\centering
\small
\caption{Descriptive statistics: number of dialogues (\#Dial) and utterances (\#Utt) are total counts, while values for the number of turns (\#Turns), preferences (\#Prefs), and recommendations (\#Recs) are reported as mean $\pm$ standard deviation. Success rate (Success) is the proportion of dialogues where participants found the target item.}
\label{tab:basic_stats}
\begin{tabular*}{\textwidth}{@{\extracolsep{\fill}}lcccccc}
\toprule
\textbf{Domain} & \textbf{\#Dial} & \textbf{\#Utt} & \textbf{\#Turns} & \textbf{\#Prefs} & \textbf{\#Recs} & \textbf{Success} \\
\midrule
Bicycle & 79 & 1521 & $9.53\pm4.71$ & $10.00\pm4.83$ & $3.19\pm1.35$ & 67.9\%\\
Digital Camera & 79 & 1687 & $10.35\pm4.42$ & $10.76\pm5.23$ & $3.42\pm1.72$ & 29.1\% \\
Laptop & 98 & 2179 & $10.00\pm4.86$ & $10.47\pm4.51$ & $3.60\pm1.27$ & 29.6\% \\
Running Shoes & 179 & 3636 & $10.07\pm5.37$ & $9.90\pm5.54$ & $3.52\pm1.43$ & 37.1\% \\
Smartwatch & 80 & 1665 & $10.18\pm3.63$ & $9.61\pm4.08$ & $3.45\pm1.19$  & 37.5\% \\
\midrule
\textbf{Total} & \textbf{515} & \textbf{10688}  & \textbf{10.21$\pm$4.81} & \textbf{10.10$\pm$5.03} & \textbf{3.46$\pm$1.41}  & \textbf{39.3\%} \\
\bottomrule
\end{tabular*}
\end{table*}

\subsection{Study Flow}

Our study consists of four sequential stages: a screening questionnaire, a domain knowledge assessment, the main task, and an exit questionnaire. This flow is designed to balance participants across domains and expertise levels while collecting both self-reported and measured indicators of domain knowledge. While the self-reported knowledge categories used in our study offer structure, we acknowledge that the boundaries between them are soft, with many participants falling in the middle of the spectrum.

\boldheading{Screening}
Participants first complete a brief self-assessment of their expertise corresponding to each of the five domains used in this work: bicycles, laptops, digital cameras, running shoes, and smartwatches. For each domain, they classify themselves into one of three categories: \emph{novice}, \emph{intermediate}, or \emph{expert}. After the screening, they are assigned to the domain with the fewest participants for their expertise level at the time of assignment. Thus, the self-assessment provides a coarse estimate of knowledge, which is used to balance participation, ensuring that each expertise level is represented approximately equally across domains.

\boldheading{Domain Knowledge Questionnaire}
Following the self-assessment, participants complete a short factual knowledge test in the domain to which they are assigned. This step provides us with the objective estimate of their true expertise. For each domain, we constructed a set of 30 factual statements using \emph{ChatGPT 5.1}, guided by principles from Item Response Theory~\cite{baker2004item,cai2016item}. In particular, the questions were designed to span a range of difficulties and to test both surface-level familiarity and deeper domain-specific knowledge. The questionnaire adopts the certainty-in-knowledge format proposed by \citet{Vidigal:2025:Political}. Each item presents a factual statement about the domain, and participants rate its correctness using one of five options: \emph{Definitely True}, \emph{Probably True}, \emph{I don’t know}, \emph{Probably False}, or \emph{Definitely False}.
This captures both accuracy and confidence, and is more informative than binary correctness.
To provide an additional validity check, human domain experts verified the suitability of the ChatGPT-generated questions.

\boldheading{Main Task}
Participants are then presented with the instructions for \RecQuest and perform the main task presented in Section~\ref{data:game}. The assigned domain determines the item pool from which each participant is randomly assigned one of three target items and its corresponding background story. Each session lasts up to fifteen minutes, with the interaction process and system behavior governed by the game mechanics and dialogue flow previously outlined.

\boldheading{Exit questionnaire}
After completing the main task, participants fill out an exit questionnaire in which they rate the quality of the interaction and provide open-ended feedback on their experience. These responses offer a subjective account of interaction satisfaction and help identify areas for improvement in both the system and the overall study design.

\subsection{Participants}

Participants were recruited from the Prolific platform,\footnote{https://www.prolific.com/} {residing in the US or Europe with English as first or primary language}
and had a $\geq 95\%$ approval rate and $\geq 1000$ previous submissions. Each participant could take part in only one task to avoid repeated exposure to the experimental setup. Participants completed the study fully online. After reading the task instructions, they proceeded through the screening questionnaire, domain knowledge test, main task, and exit questionnaire. Sessions lasted for 18 minutes on average. Participants received an hourly wage of £8, in line with the platform's fair-wage guidelines, with a £1 bonus awarded on successful identification of the target items.

\subsection{Generating Game States}
\label{sec:data_collection:generating_game_states}

\boldheading{Item Collection}
To operationalize the \RecQuest protocol, we instantiated item collections and target scenarios. From the Amazon dataset, we extracted all items belonging to categories related to the five domains. The raw collections ranged from 1,000 to 200,000 items per domain and included many near-duplicates differing only in brand or minor variations. To clean this data and maintain sufficient item differentiation, we narrowed each domain to a curated collection of 50 items. We achieved this by embedding item descriptions and applying k-means clustering to the embeddings, selecting the items nearest to the cluster centroids. The same embedding representations were later used for retrieval during gameplay.

\boldheading{Target Items}
To create the interactive scenarios, we prepared a total of 30 background stories. For each domain, we selected three unique items from the curated collection and generated two versions of a story for each item: a \emph{long} version and a \emph{short} version. Each story was designed to contain a fixed number of implicit feature constraints—ten for the long version and five for the short—sufficient to uniquely identify the target item. Using an LLM, we converted these structured constraints into natural narratives without explicitly mentioning the corresponding item features.

\boldheading{LLM Usage}
The CRS pipeline relies on multiple calls to an LLM, with different prompts and inputs for each component of the system. For all LLM calls in this study, we employed \emph{gpt-4.1-mini-2025-04-14}, which early tests indicated offered a good balance between speed and output quality. All prompts used in the LLM-based components of the system can be found in the GitHub repository.

\subsection{Analysis}

We now address \emph{\textbf{RQ1}}, assessing whether the \RecQuest protocol achieves its primary goal: eliciting rich signals of domain knowledge through natural, engaging conversation. To answer this, we evaluate the \RecQuest protocol through two complementary lenses: a quantitative analysis of dialogue statistics to verify that the protocol successfully induces the iterative constraint refinement necessary to surface knowledge signals, and a qualitative analysis of participant feedback to validate that the resulting interactions were perceived as natural and engaging.

\boldheading{Quantitative Analysis: Interaction Depth}
The aggregate dialogue statistics provide strong evidence that the protocol elicits rich interaction signals. Overall, we collected a total of 515 high-quality conversations totaling over 10,000 utterances across five domains. A statistical overview of the collected data is presented in Table~\ref{tab:basic_stats}. Crucially, dialogues average $10.21$ turns and contain $10.10$ preference statements. This density of preference expression, combined with an average of $3.46$ recommendations per session, indicates that participants did not merely search for items, but engaged in iterative refinement over multiple exchanges. This confirms that the protocol successfully drives the articulation of detailed constraints, which serve as the proxies for domain knowledge in our analysis.

\boldheading{Qualitative Analysis: User Experience}
To validate the ``engaging and natural'' aspect of RQ1, we examine the feedback from the post-task questionnaire. Responses were overall positive (mean rating of 3.89/5.0), with comments  highlighting the educational and conversational nature of the system: \emph{P13: ``I appreciated how the chatbot patiently asked clarifying questions to understand my priorities, which made the recommendations feel personalized and relevant. The detailed comparisons between different e-bikes, including specific features like battery capacity, motor type, braking system, and handling, helped me make an informed decision},'' and \emph{P66: ``The chatbot is very friendly and human like. I enjoyed every bit}.'' 

Negative comments primarily focused on the study constraints, such as the limited collection (e.g., \emph{P396: ``It could do with offering more [laptop] models.}'') or the interaction constraint that prohibits information dumps (e.g., \emph{P502: ``was quite frustrating because it didn't take into account all of my wants all at once.}''). However, this latter frustration indirectly validates the protocol design: it confirms that the system successfully prevented keyword-search behaviors, forcing users to explore the domain through dialogue.
Overall, the positive feedback on personalization and the fluid interaction confirms our protocol's effectiveness in creating an engaging experience at scale. 

\section{Estimating User Domain Knowledge}

In this section, we begin by comparing participants' self‑reported expertise with their actual performance on the domain‑knowledge questionnaires (\textbf{RQ2}). To move past the well-documented unreliability of self-assessments~\cite{dang2020self,cole2010self}, we investigate multiple strategies for grouping participants by objective scores on the knowledge questionnaires. We then examine behavioral patterns that distinguish experts from novices in real interaction settings (\textbf{RQ3}). Building on these validated ground‑truth labels, we finally evaluate whether LLMs can infer user knowledge directly from dialogue transcripts (\textbf{RQ4}). Our approach uses both zero‑shot and few‑shot prompting to estimate overall knowledge levels, circumventing dependence on large training datasets while leveraging the natural language understanding capabilities of modern pre-trained models.

\subsection{Methods}

We estimate user domain knowledge directly from conversation transcripts using LLMs in zero-shot and few-shot settings. All methods share a unified prompt structure, shown in Fig.~\ref{fig:prompt}, comprising a fixed task description, a three-level expertise definition, placeholders for the domain, conversation history, optional few-shot examples, and a constrained JSON output format, to ensure consistency across settings. We evaluate three prompt variants. The \emph{holistic} variant conditions on the full conversation history and produces a single overall knowledge estimate based on all user turns. The \emph{evidence-based} variant also uses the full conversation but first instructs the model to extract domain‑specific phrases introduced by the user and to ground its prediction explicitly in this extracted evidence. The \emph{incremental} variant operates turn-by-turn: given the conversation history, a previous estimate, and the latest user utterance, the model updates the predicted knowledge level only when new evidence appears, with the restriction that the estimate may increase or remain stable but not decrease---a pragmatic design choice that prioritizes stability over perfect Bayesian updating. Each variant is evaluated in both zero-shot and few-shot settings, with few-shot prompts providing three domain-matched examples (one for each label). We additionally train a supervised fine‑tuned model using the same prompt structure and annotated conversations.

\begin{figure}[t]
\begin{promptbox}
    \sffamily\footnotesize
        
    Role:
You are an expert system for evaluating a user’s domain knowledge in 
**{domain}** domain based on their conversation with a recommender system.\\

Task: \\
Estimate the user’s expertise level based on their utterances. \\

Knowledge Levels: \\
Level 1 (Novice): No independent technical knowledge. \\
Level 2 (Intermediate): Some correct use of domain terminology. \\
Level 3 (Expert): Demonstrates knowledge and independent use of 
    domain concepts. \\

Input Handling:\\
\textcolor{vargreen}{$\langle$if mode = holistic$\rangle$}\\
\phantom{xxxx}You are given the entire conversation history.\\
\phantom{xxxx}Assess all user turns jointly and produce a single overall estimate.\\
\textcolor{vargreen}{$\langle$elif mode = phrase-identification$\rangle$}\\
\phantom{xxxx}You are given the entire conversation history.\\
\phantom{xxxx}First, extract domain-specific phrases that were introduced by the user, \\
\phantom{xxxx}then base the knowledge estimate on those phrases. \\
\textcolor{vargreen}{$\langle$elif mode = incremental$\rangle$}\\
\phantom{xxxx}You are given the conversation history, a previous knowledge estimate,\\
\phantom{xxxx}and the latest user utterance.\\
\phantom{xxxx}Update the estimate only if the latest turn provides new evidence.\\
\phantom{xxxx}The level may increase or remain the same, but never decrease.\\
\textcolor{vargreen}{$\langle$end if$\rangle$}\\

Input Data:\\
\textcolor{varblue}{[conversation history]}\\

\textcolor{vargreen}{$\langle$if mode = incremental$\rangle$}\\
\phantom{xxxx}Previous estimate: \textcolor{varblue}{[previous estimate]} \\
\phantom{xxxx}Latest user utterance: \textcolor{varblue}{[latest user utterance]} \\
\textcolor{vargreen}{$\langle$end if$\rangle$}\\

\textcolor{vargreen}{$\langle$if fewshot$\rangle$}\\
\phantom{xxxx}\textcolor{varblue}{[Examples]} \\
\textcolor{vargreen}{$\langle$end if$\rangle$}\\

Format:\\
$\{$\\
\textcolor{vargreen}{$\langle$if mode = phrase-identification$\rangle$}\\
\phantom{xxxx}"user\_initiated\_phrases": [...], \\
\textcolor{vargreen}{$\langle$end if$\rangle$}\\
\phantom{xxxx}"knowledge\_level": <1, 2, or 3>, \\
\phantom{xxxx}"reasoning": "<brief explanation>" \\
$\}$\\

All responses must be valid JSON. \\
Output:
\end{promptbox}
    \caption{LLM prompt used for knowledge estimation. Placeholders for variables are in \textcolor{varblue}{[square brackets]}, while conditional logic is marked by \textcolor{vargreen}{$\langle$if$\rangle$}...\textcolor{vargreen}{$\langle$end if$\rangle$}.}
    \label{fig:prompt}
\end{figure}

\subsection{Experimental Details}

We evaluate two off-the-shelf LLMs for zero-shot and few-shot prompting: \emph{gpt-4.1-mini-2025-04-14}---also used during data collection---and \emph{Meta-Llama-3.3-70B}, an open-weight model. Supervised fine-tuning is performed on a 4bit quantized \emph{Meta-Llama-3-8B-Instruct} model with QLoRA adapters and an encoder-only \emph{ModernBERT-large} model suitable for a classification task, using 310 labeled conversations, with 205 held out for evaluation. Both models were trained for 10 epochs and $2e-5$ learning rate. All models are evaluated using identical prompts, and expertise definitions.
\section{Experimental Results}
\label{sec:results}

Here, we present our findings corresponding  \textbf{RQ2}, \textbf{RQ3}, and \textbf{RQ4}. 

\subsection{RQ2: Establishing Ground Truth} 
\label{sec:results:rq2}

To characterize user behavior accurately, we first require a reliable method for labeling expertise. We
compare the reliability of subjective self-assessments against objective knowledge scores.

\begin{figure}[t]
  \centering
    \includegraphics[width=1\linewidth]{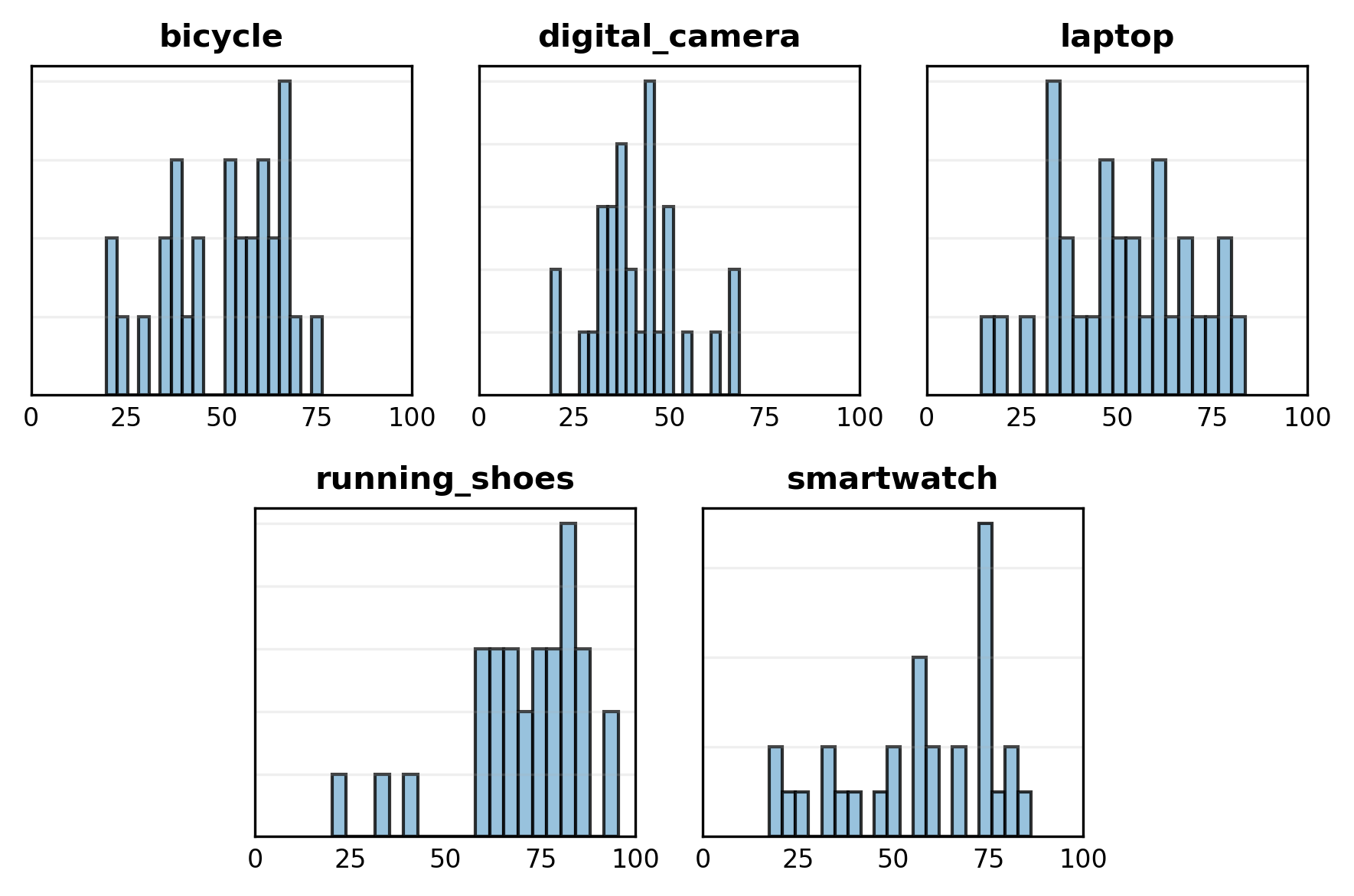}
  \caption{Distribution of questions by fraction of participants who correctly answered them by domain.
  }
  \label{fig:knowledge_test}
\end{figure}

\begin{figure}[t]
  \centering
    \includegraphics[width=0.95\linewidth]{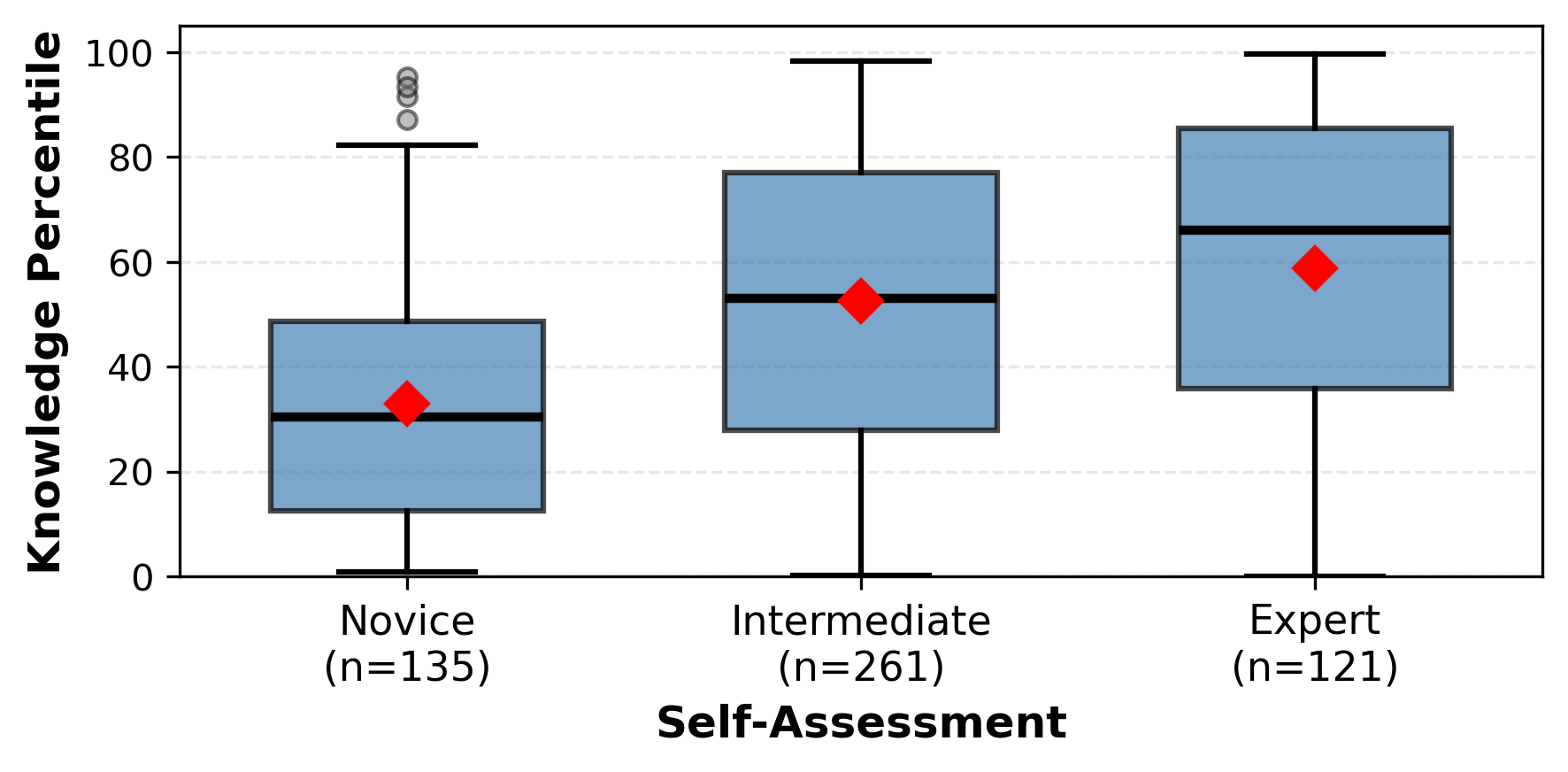}
  \caption{Relationship between objective and self-assessed knowledge.}
  \label{fig:knowledge}
\end{figure}

Figure~\ref{fig:knowledge_test} presents the distribution of item difficulties per domain, measured by the proportion of participants answering each question correctly. 
The spread is well‑balanced: no item was prohibitively difficult (each was solved by at least 15\% of participants), very few were trivial (solved by more than 90\%), and most fell between 40\% and 80\%. This distribution indicates that the questionnaire is reasonably well‑suited for distinguishing between different knowledge levels in our study setting.

We then examined how participants' self‑reported expertise relates to their objective knowledge scores (Figure~\ref{fig:knowledge}). The correlation is weak, with notable discrepancies---particularly among participants who identified as Intermediate or Expert yet performed poorly on the objective test. 
To empirically determine which metric serves as a more reliable ground truth, we compared task success rates under both labeling schemas. Intuitively, higher domain knowledge should facilitate the search process and lead to better outcomes. Indeed, when using objective scoring, Experts outperformed Novices (40.4\% vs. 35.9\%). However, when grouping users by self-assessment, this trend paradoxically reversed: self-identified Novices achieved a higher success rate (44.4\%) than self-identified Experts (36.4\%). This inversion---where perceived expertise negatively correlates with actual task performance---provides compelling empirical evidence for the Dunning–Kruger phenomenon~\cite{kruger1999unskilled}, where overconfidence masks a lack of competence.

This confirms that self‑assessment is a noisy and misleading signal. 
In contrast, objective scores offer a more granular and verifiable basis for labeling. Consequently, for the remainder of our analysis (RQ3 and RQ4), we use objective score percentiles as an operationalized ground-truth proxy, defining \textbf{Novices} ($\leq$ 20th percentile) and \textbf{Experts} ($\geq$ 80th percentile).

\subsection{RQ3: Behavioral and Linguistic Traces}
\label{sec:results:rq3}

Having established a robust ground truth, we examine how domain knowledge manifests within the conversation. Our goal is to determine whether experts and novices exhibit distinct behavioral patterns that are consistent enough to separate the two groups in practice. We analyze these interaction traces across three dimensions: dialogue flow, task outcomes, and linguistic specificity.

\begin{figure}[t]
  \centering
    \includegraphics[width=0.85\linewidth]{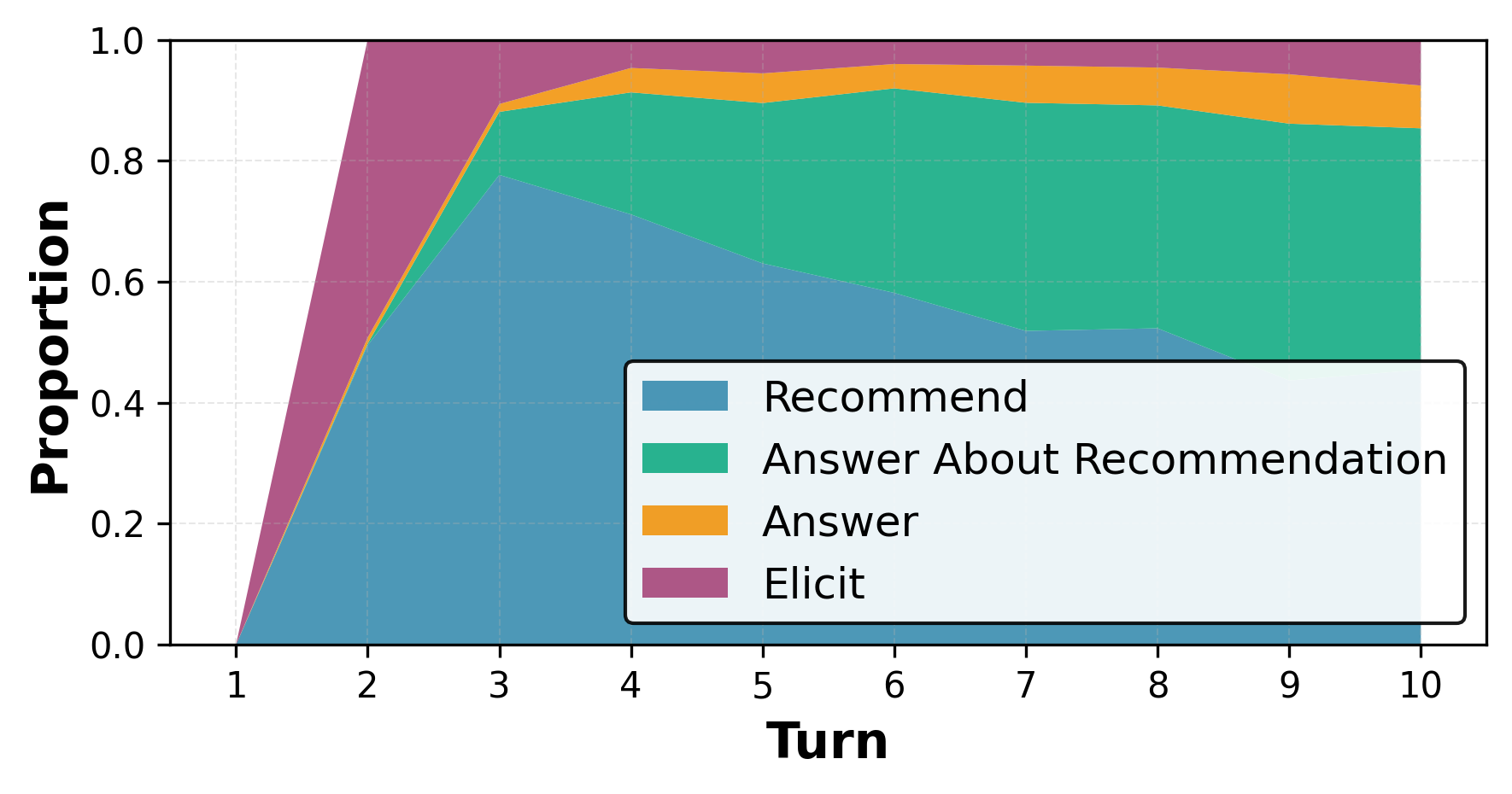}
  \caption{Dialogue intent progression.}
  \label{fig:intents}
\end{figure}

\boldheading{Dialogue Patterns and Task Outcomes} The dialogue intent progression, (Fig.~\ref{fig:intents} shows that \emph{Recommendation} dominates early interactions, peaking around the third turn before gradually declining, as \emph{Answer About Recommendation} intents increase over time. 
Although all groups follow this general trajectory, we observe subtle differences. Experts receive slightly more recommendations, consistent with their more frequent preference updates and refinements. Novices, in contrast, ask more clarifying questions, indicating a greater need for guidance. 
Beyond the flow of conversation, we also analyze \emph{success rate}: the percentage of participants who correctly identified the target item. Success varied substantially across domains, from 29.1\% (running shoes) to 67.8\% (bicycles), with an overall rate of 39.3\%. Experts (40.4\%) outperformed novices (35.9\%), suggesting that domain knowledge helps users steer the system more effectively.

\begin{figure}[t]
  \centering
        \includegraphics[width=0.85\linewidth]{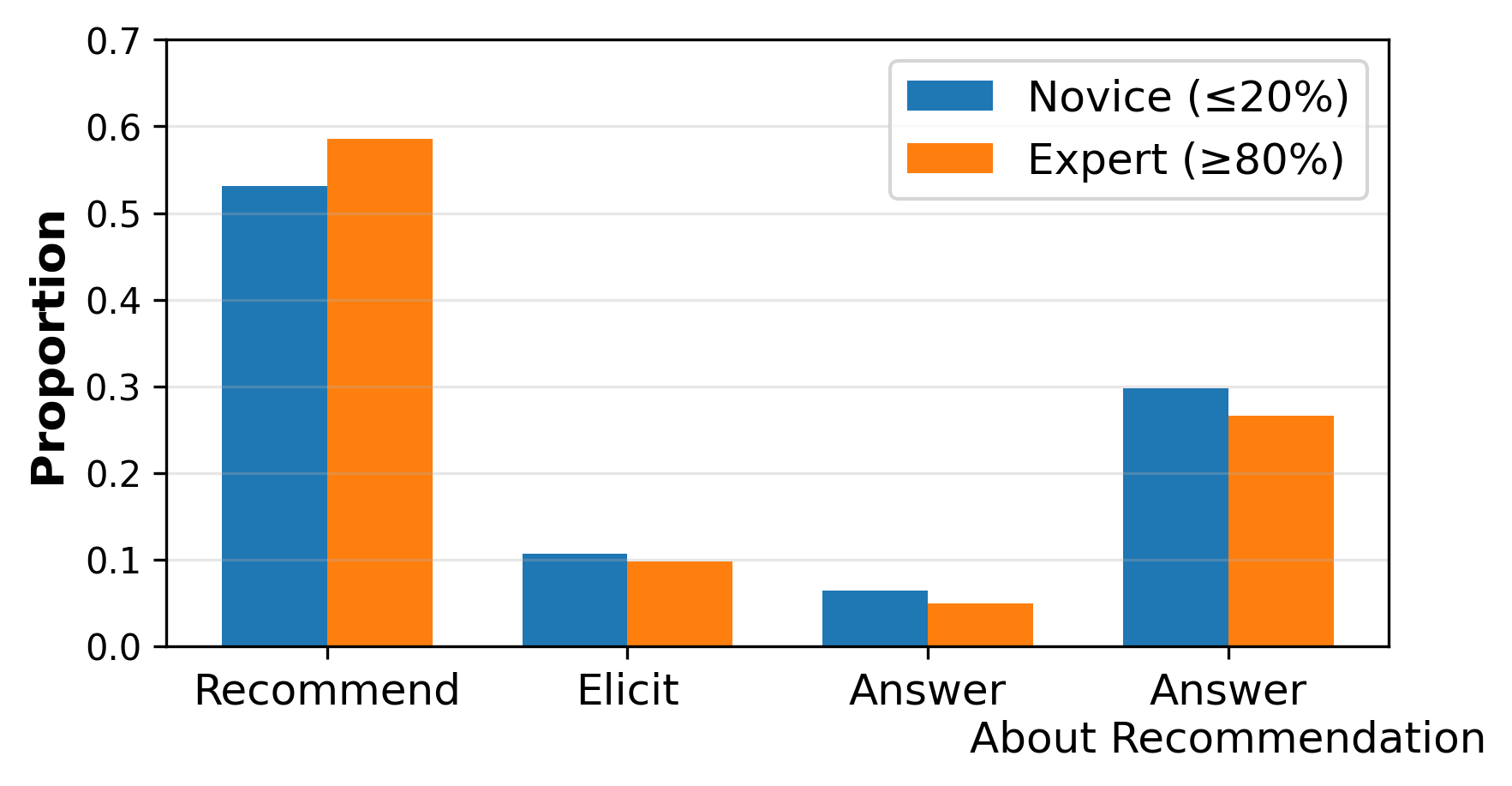}
  \caption{Distribution of dialogue acts.}
  \label{fig:actions}
\end{figure}

\boldheading{Linguistic Markers of Expertise} To further answer RQ3, we first analyze basic linguistic features, such as the number of turns and utterance length. Experts engaged in slightly longer dialogues (10.68 messages on average) and used longer utterances (17.3 words) than novices (9.87 messages, 14.0 words).

Next, we compute domain-specific TF-IDF scores to quantify differences in keyword usage between the novice and expert groups. For each domain, we construct joint TF-IDF models over all documents to ensure a shared vocabulary and comparable scaling. Then, we average TF-IDF values within each group. The resulting normalized differences indicate how strongly each keyword is associated with Experts versus Novices (ranging from -1 to +1).

\begin{table}[t]
\centering
\caption{Discriminative keywords by domain and expertise.}
\scriptsize
\begin{tabularx}{\linewidth}{l|X|X}
\toprule
\textbf{Domain} & \textbf{Novice} ($\leq 20th$ percentile) & \textbf{Expert} ($\geq 80th$ percentile) \\ 
\midrule
Bicycle & recommend, local, new, need, stay & battery, range, assist, brake, gravel, trail, motor \\
Digital Camera & camera, bird, look, zoom, stability & lens, FujiFilm, autofocus, fast, travel \\ 
Laptop & best, game, work, time, speed, smooth & power, performance, light, windows, lenovo \\ 
Running Shoes & comfort, look, breathable & trail, Salomon, grip, protect \\
Smartwatch & smartwatch, good, brightness & battery life, charging, appearance  \\ 
\bottomrule
\end{tabularx}
\label{tab:discriminative_keywords}
\end{table}

Table~\ref{tab:discriminative_keywords} lists representative keywords highlighting the contrasts in keyword usage. Across domains, novices tend to use more general or evaluative terms (e.g., good, recommend, comfort),  reflecting subjective impressions or purchase intentions. Experts, in contrast, rely more on technical or domain-specific vocabulary (e.g., brake, autofocus, performance), emphasizing specifications and functional attributes. These patterns suggest that Experts approach the CRS with a more analytical, comparison-oriented communication style, while novices rely on broader, experience-based descriptions.

This TF-IDF analysis provides preliminary evidence for a clear hypothesis: \emph{Experts tend to communicate through specific product attributes, while novices tend to communicate through intended usage or scenarios}.
To test this hypothesis systematically, we designed an LLM-based annotation pipeline. We few-shot-prompted \emph{gpt-4.1-mini-2025-04-14} to analyze each user utterance and extract phrases corresponding to two distinct categories: (i) \emph{Specific Attributes}: Mentions of technical features, components, or specifications (e.g., ``disc brakes'').
(ii) \emph{Intended Usage}: Descriptions of goals, contexts, or scenarios (e.g., ``for daily rides to work''). This annotation moves beyond simple keyword counts and enables per-utterance quantification of communication style.
The results strongly support our hypothesis, particularly at the conversation level. We found that Experts mentioned significantly more specific attributes on average (9.01 per conversation) than Novices (6.80). While Experts also provided more usage-based context (5.38 vs. 4.59), their overall linguistic preference still leaned more heavily toward attributes (62.6\% of expressions) compared to Novices (59.8\%).

\subsection{RQ4: Automatic Knowledge Estimation}

\begin{table}[t]
\centering
\small
\caption{Results of automatic knowledge estimation.}
\label{tab:overall_macro_performance}
\setlength{\tabcolsep}{8pt}
\begin{tabular}{clcc}
\hline
 & \textbf{Model}
 & \textbf{Accuracy}
 & \textbf{Macro-F1} \\
\hline

\multirow{4}{*}{Holistic}
& GPT-4.1 (zero-shot)  & 0.305 & 0.296 \\
& GPT-4.1 (few-shot)   & 0.378 & \textbf{0.363} \\
& LLaMA 3.3 (zero-shot)& 0.354 & 0.354 \\
& LLaMA 3.3 (few-shot) & \textbf{0.490} & 0.357 \\

\hline
\multirow{4}{*}{Incremental}
& GPT-4.1 (zero-shot)  & 0.359 & 0.332 \\
& GPT-4.1 (few-shot)   & 0.349 & 0.347 \\
& LLaMA 3.3 (zero-shot)& 0.364 & 0.347 \\
& LLaMA 3.3 (few-shot) & \textbf{0.490} & \textbf{0.356} \\

\hline
\multirow{4}{*}{Evidence-based}
& GPT-4.1 (zero-shot)  & 0.339 & 0.321 \\
& GPT-4.1 (few-shot)   & 0.417 & 0.379 \\
& LLaMA 3.3 (zero-shot)& \textbf{0.417} & \textbf{0.382} \\
& LLaMA 3.3 (few-shot) & 0.213 & 0.154 \\

\hline
\multirow{2}{*}{Supervised} & Llama 3-8B  & 0.519 & 0.286 \\
 & ModernBERT  & \textbf{0.524} & \textbf{0.353} \\

\hline
\end{tabular}
\end{table}

In this section, we address RQ4 by examining how effectively large language models can estimate user domain knowledge from conversational interactions. 

Table~\ref{tab:overall_macro_performance} reports classification performance using both accuracy and macro-averaged metrics to capture global correctness and class-level balance. Zero-shot LLaMA consistently yields strong macro-F1 scores, indicating robust alignment across all expertise levels. We observe contrasting few-shot effects: adding in-context examples consistently improves GPT-4.1 but degrades LLaMA 3.3, particularly in the evidence-based setting where performance drops substantially. Among strategies, incremental estimation proves most effective. While the fine-tuned ModernBERT achieves the highest overall accuracy ($0.524$), it struggles with underrepresented categories, showing higher variability than the prompted LLMs. Ultimately, these moderate results highlight the inherent difficulty of the task: expertise cues are rarely explicit in single utterances, but rather distributed indirectly across multiple conversational turns.

\begin{figure}[t]
  \centering
    \includegraphics[width=\linewidth]{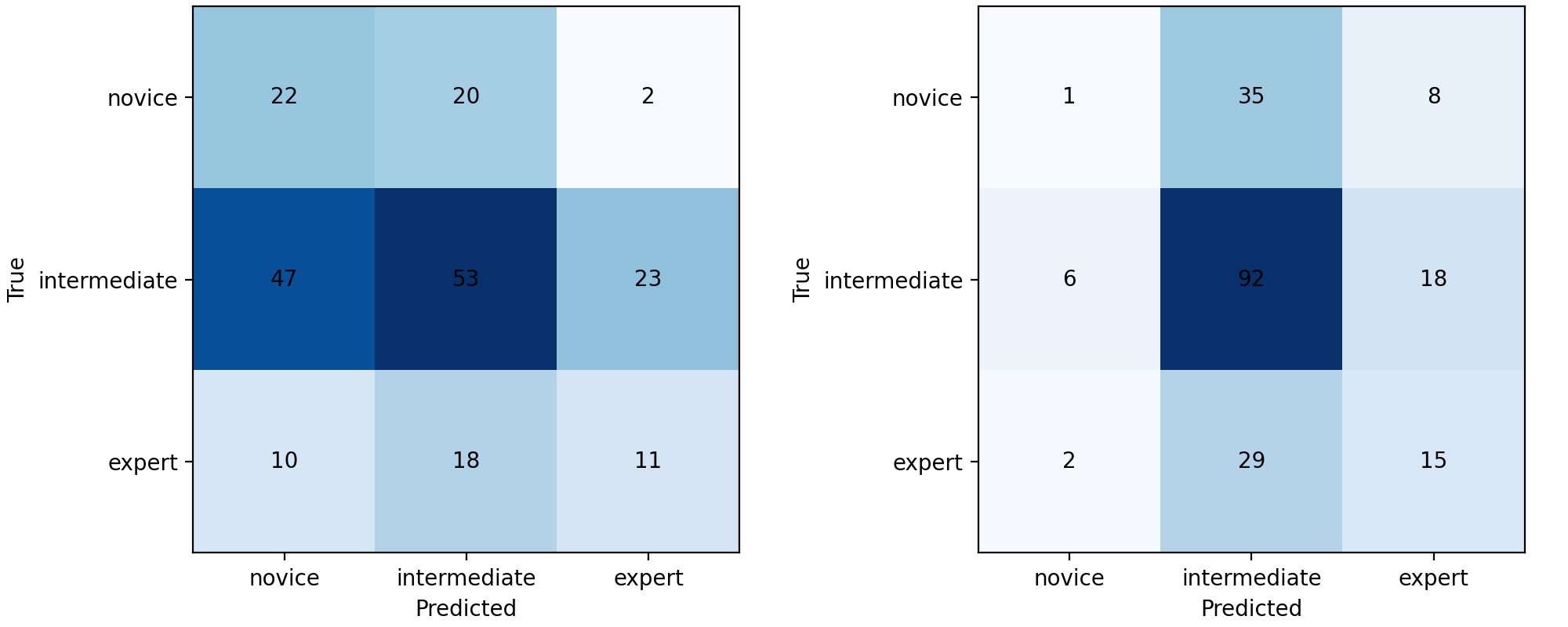}
  \caption{Confusion matrix comparison of best performing prompting model (Evidence-based, zero-shot LLama) left, and ModernBERT on the right.}
  \label{fig:confusion}
\end{figure}

To further understand the challenges in automatic knowledge estimation, Figure~\ref{fig:confusion} contrasts confusion matrices for two of the best performing methods: zero-shot LLaMA and supervised ModernBERT. LLaMA shows distributed predictions with errors mostly confined to adjacent classes. Conversely, ModernBERT strongly favors the Intermediate class, maximizing overall accuracy at the expense of recall for Novices and Experts. This distinction explains the observed trade-off: ModernBERT over-centralizes predictions to optimize accuracy, whereas LLaMA maintains a better balance across the spectrum, resulting in higher macro-F1.

\section{Conclusion and Future Directions}
\label{sec:concl}

In this work, we introduce the novel task of estimating user domain knowledge from dialogue interactions, a crucial first step that enables the creation of adaptive conversational recommender systems. 
Our primary contribution is the design and validation of \RecQuest, a gamified data collection protocol that successfully elicits natural, knowledge-rich interactions. By systematically creating information gaps between users and the system, \RecQuest provokes the linguistic and behavioral traces necessary for studying domain knowledge in a CRS context.
A key finding of our study is the necessity of objective knowledge assessment. Our comparison of self-reports versus objective scores revealed significant calibration biases, indicating that self-assessment is a noisy signal of expertise in this setting. 
Future research in this area must therefore rely on verified competence rather than user confidence.
Finally, we established baselines for the automatic estimation of user knowledge. Our initial LLM-based experiments indicate that while broad distinctions between novices and experts are possible, fine-grained estimation remains an inherently difficult task. The subtle and evolving nature of knowledge signals in short conversational turns poses a significant challenge for current models, and the current results should thus be interpreted as initial evidence of task feasibility rather than robust practical performance.

Our protocol and dataset provide a foundation for studying this challenge and supporting future CRS that adapt elicitation and explanation strategies to users with different levels of understanding. More broadly, our results suggest that robust estimation will likely require additional data, and that the protocol may need refinement depending on the target application domain and conversational context.
Finally, an inherent limitation of our current study is the assumption that differences in conversational behavior stem primarily from domain expertise. We recognize that real-world interactions are heavily influenced by confounding factors like individual verbosity, communication style, language proficiency, and personality traits. Isolating true knowledge signals from these individual characteristics remains an important challenge for future work.

\begin{acks}
    An unrestricted gift from Google partially supported this research.
\end{acks}

\bibliographystyle{ACM-Reference-Format}
\balance
\bibliography{ictir2026-knowledge.bib}

@inproceedings{Bernard:2023:SIGIR,
  title = {MG-ShopDial: A Multi-Goal Conversational Dataset for e-Commerce},
  booktitle = {Proceedings of the 46th International ACM SIGIR Conference on Research and Development in Information Retrieval},
  author = {Bernard, Nolwenn and Balog, Krisztian},
  year = {2023},
  series = {SIGIR '23},
  pages = {2775--2785}
}

@article{Vidigal:2025:Political,
  title={Measuring Belief Certainty in Political Knowledge},
  author={Vidigal, Robert},
  journal={Political Behavior},
  volume={47},
  number={2},
  pages={529--551},
  year={2025},
}

@book{baker2004item,
  title={Item response theory: Parameter estimation techniques},
  author={Baker, Frank B and Kim, Seock-Ho},
  year={2004},
  publisher={CRC press}
}

@article{cai2016item,
  title={Item response theory},
  author={Cai, Li and Choi, Kilchan and Hansen, Mark and Harrell, Lauren},
  journal={Annual Review of Statistics and Its Application},
  volume={3},
  number={1},
  pages={297--321},
  year={2016}
}

@article{kruger1999unskilled,
  title={Unskilled and unaware of it: how difficulties in recognizing one's own incompetence lead to inflated self-assessments.},
  author={Kruger, Justin and Dunning, David},
  journal={Journal of personality and social psychology},
  volume={77},
  number={6},
  pages={1121--1134},
  year={1999}
}

@article{cole2010self,
  title={Are self-assessments reliable indicators of topic knowledge?},
  author={Cole, Michael J and Zhang, Xiangmin and Liu, Jinging and Liu, Chang and Belkin, Nicholas J and Bierig, Ralf and Gwizdka, Jacek},
  journal={Proceedings of the American Society for Information Science and Technology},
  volume={47},
  number={1},
  pages={1--10},
  year={2010}
}

@article{dang2020self,
  title={Why are self-report and behavioral measures weakly correlated?},
  author={Dang, Junhua and King, Kevin M and Inzlicht, Michael},
  journal={Trends in cognitive sciences},
  volume={24},
  number={4},
  pages={267--269},
  year={2020}
}

@inproceedings{collins2016assessing,
  title={Assessing learning outcomes in web search: A comparison of tasks and query strategies},
  author={Collins-Thompson, Kevyn and Rieh, Soo Young and Haynes, Carl C and Syed, Rohail},
  booktitle={Proceedings of the 2016 ACM on conference on human information interaction and retrieval},
  pages={163--172},
  year={2016},
  series={CHIIR '16}
}

@inproceedings{gadiraju2018analyzing,
  title={Analyzing knowledge gain of users in informational search sessions on the web},
  author={Gadiraju, Ujwal and Yu, Ran and Dietze, Stefan and Holtz, Peter},
  booktitle={Proceedings of the 2018 conference on human information interaction \& retrieval},
  pages={2--11},
  year={2018},
  series={CHIIR '18}
}

@inproceedings{balayn2022ready,
  title={Ready player one! eliciting diverse knowledge using a configurable game},
  author={Balayn, Agathe and He, Gaole and Hu, Andrea and Yang, Jie and Gadiraju, Ujwal},
  booktitle={Proceedings of the ACM Web Conference 2022},
  pages={1709--1719},
  year={2022},
  series={WWW '22}
}

@incollection{Brusilovsky:2007:Springer,
  author = {Brusilovsky, Peter and Mill{\'a}n, Eva},
  title = {User Models for Adaptive Hypermedia and Adaptive Educational Systems},
  booktitle = {The Adaptive Web: Methods and Strategies of Web Personalization},
  pages = {3--53},
  publisher = {Springer},
  year = {2007}
}

@inproceedings{Budzianowski:2018:EMNLP,
  title = {MultiWOZ - A Large-Scale Multi-Domain Wizard-of-Oz Dataset for Task-Oriented Dialogue Modelling},
  booktitle = {Proceedings of the 2018 Conference on Empirical Methods in Natural Language Processing},
  author = {Budzianowski, Pawe{\l} and Wen, Tsung-Hsien and Tseng, Bo-Hsiang and Casanueva, I{\~n}igo and Ultes, Stefan and Ramadan, Osman and Ga{\v s}i{\'c}, Milica},
  year = {2018},
  series = {EMNLP '18},
  pages = {5016--5026}
}

@inproceedings{Joko:2024:SIGIR,
  title = {Doing Personal LAPS: LLM-Augmented Dialogue Construction for Personalized Multi-Session Conversational Search},
  booktitle = {Proceedings of the 47th International ACM SIGIR Conference on Research and Development in Information Retrieval},
  author = {Joko, Hideaki and Chatterjee, Shubham and Ramsay, Andrew and De Vries, Arjen P. and Dalton, Jeff and Hasibi, Faegheh},
  year = {2024},
  series = {SIGIR '24},
  pages = {796--806}
}

@inproceedings{Rastogi:2020:AAAI,
  title = {Towards Scalable Multi-Domain Conversational Agents: The Schema-Guided Dialogue Dataset},
  author = {Rastogi, Abhinav and Zang, Xiaoxue and Sunkara, Srinivas and Gupta, Raghav and Khaitan, Pranav},
  booktitle={Proceedings of the AAAI conference on artificial intelligence},
  year = {2020},
  pages = {8689--8696},
  series={AAAI '20}
}

@inproceedings{Radlinski:2019:SIGDIAL,
  title = {Coached Conversational Preference Elicitation: A Case Study in Understanding Movie Preferences},
  booktitle = {Proceedings of the 20th Annual SIGdial Meeting on Discourse and Dialogue},
  author = {Radlinski, Filip and Balog, Krisztian and Byrne, Bill and Krishnamoorthi, Karthik},
  year = {2019},
  series = {SIGDIAL '19},
  pages = {353--360}
}

@inproceedings{Shah:2018:NAACL,
  title = {Bootstrapping a Neural Conversational Agent with Dialogue Self-Play, Crowdsourcing and On-Line Reinforcement Learning},
  booktitle = {Proceedings of the 2018 Conference of the North American Chapter of the Association for Computational Linguistics: Human Language Technologies, Volume 3 (Industry Papers)},
  author = {Shah, Pararth and {Hakkani-T{\"u}r}, Dilek and Liu, Bing and T{\"u}r, Gokhan},
  year = {2018},
  series = {NAACL '18},
  pages = {41--51}
}

@inproceedings{Kostric:2025:UMAP,
  author =    {Kostric, Ivica and Balog, Krisztian and Gadiraju, Ujwal},
  title =     {Should We Tailor the Talk? Understanding the Impact of Conversational Styles on Preference Elicitation in Conversational Recommender Systems},
  booktitle = {Proceedings of the 33rd ACM Conference on User Modeling, Adaptation and Personalization},
  pages = {164--173},
  series =    {UMAP '25},
  year =      {2025},
}

@article{Pramod:2022:Expert,
  title = {Conversational recommender systems techniques, tools, acceptance, and adoption: A state of the art review},
  author = {Pramod, Dhanya and Bafna, Prafulla},
  year = {2022},
  journal = {Expert Systems with Applications},
  volume = {203},
  pages = {117539}
}

@article{Kostric:2024:TORS,
  title = {Generating Usage-related Questions for Preference Elicitation in Conversational Recommender Systems},
  author = {Kostric, Ivica and Balog, Krisztian and Radlinski, Filip},
  year = {2024},
  journal = {ACM Transactions on Recommender Systems},
  volume = {2},
  number = {2},
  pages = {1--24}
}

@article{Jannach:2022:ACM,
  title = {A Survey on Conversational Recommender Systems},
  author = {Jannach, Dietmar and Manzoor, Ahtsham and Cai, Wanling and Chen, Li},
  year = {2022},
  journal = {ACM Computing Surveys},
  volume = {54},
  number = {5},
  pages = {1--36}
}

@inproceedings{Lin:2023:WWW,
  title = {Enhancing User Personalization in Conversational Recommenders},
  booktitle = {Proceedings of the ACM Web Conference 2023},
  author = {Lin, Allen and Zhu, Ziwei and Wang, Jianling and Caverlee, James},
  year = {2023},
  series = {WWW '23},
  pages = {770--778}
}

@article{Zhang:2025:ACM,
  title = {Vague Preference Policy Learning for Conversational Recommendation},
  author = {Zhang, Gangyi and Gao, Chongming and Lei, Wenqiang and Guo, Xiaojie and Li, Shijun and Chen, Hongshen and Ding, Zhuozhi and Xu, Sulong and Wu, Lingfei},
  year = {2025},
  journal = {ACM Transactions on Information Systems},
    volume={43},
  number={3},
  pages={1--27}
}

@inproceedings{Zhao:2022:WWW,
  title = {Knowledge-aware Conversational Preference Elicitation with Bandit Feedback},
  booktitle = {Proceedings of the ACM Web Conference 2022},
  author = {Zhao, Canzhe and Yu, Tong and Xie, Zhihui and Li, Shuai},
  year = {2022},
  series = {WWW '22},
  pages = {483--492}
}

@inproceedings{Bernard:2024:WSDM,
  title = {IAI MovieBot 2.0: An Enhanced Research Platform with Trainable Neural Components and Transparent User Modeling},
  booktitle = {Proceedings of the 17th ACM International Conference on Web Search and Data Mining},
  author = {Bernard, Nolwenn and Kostric, Ivica and Balog, Krisztian},
  year = {2024},
  series = {WSDM '24},
  pages = {1042--1045}
}

@inproceedings{Lei:2020:WSDMa,
  title = {Estimation-Action-Reflection: Towards Deep Interaction Between Conversational and Recommender Systems},
  booktitle = {Proceedings of the 13th International Conference on Web Search and Data Mining},
  author = {Lei, Wenqiang and He, Xiangnan and Miao, Yisong and Wu, Qingyun and Hong, Richang and Kan, Min-Yen and Chua, Tat-Seng},
  year = {2020},
  series = {WSDM '20},
  pages = {304--312}
}

@inproceedings{Christakopoulou:2016:KDD,
  title = {Towards conversational recommender systems},
  booktitle = {Proceedings of the 22nd ACM SIGKDD international conference on knowledge discovery and data mining},
  author = {Christakopoulou, Konstantina and Radlinski, Filip and Hofmann, Katja},
  year = {2016},
  series = {KDD '16},
  pages = {815--824}
}

@inproceedings{Zhou:2020:COLING,
  title = {Towards Topic-Guided Conversational Recommender System},
  author = {Zhou, Kun and Zhou, Yuanhang and Zhao, Wayne Xin and Wang, Xiaoke and Wen, Ji-Rong},
  booktitle = {Proceedings of the 28th International Conference on Computational Linguistics},
  series = {COLING '20},
  pages = {4128--4139},
  year = {2020}
}

@inproceedings{Li:2018:NIPS,
  title = {Towards Deep Conversational Recommendations},
  booktitle = {Advances in Neural Information Processing Systems},
  author = {Li, Raymond and Ebrahimi Kahou, Samira and Schulz, Hannes and Michalski, Vincent and Charlin, Laurent and Pal, Chris},
  year = {2018},
  series = {NIPS '18},
   pages = {9748--9758},
    volume = {31}
}

@inproceedings{Kang:2019:EMNLP,
  title = {Recommendation as a Communication Game: Self-Supervised Bot-Play for Goal-oriented Dialogue},
  booktitle = {Proceedings of the 2019 Conference on Empirical Methods in Natural Language Processing and the 9th International Joint Conference on Natural Language Processing},
  author = {Kang, Dongyeop and Balakrishnan, Anusha and Shah, Pararth and Crook, Paul and Boureau, Y-Lan and Weston, Jason},
  year = 2019,
  series = {EMNLP-IJCNLP '19},
    pages={1951--1961}
}

@inproceedings{Hayati:2020:EMNLP,
  title = {INSPIRED: Toward Sociable Recommendation Dialog Systems},
  booktitle = {Proceedings of the 2020 Conference on Empirical Methods in Natural Language Processing},
  author = {Hayati, Shirley Anugrah and Kang, Dongyeop and Zhu, Qingxiaoyang and Shi, Weiyan and Yu, Zhou},
  year = 2020,
  series = {EMNLP '20},
  pages = {8142--8152}
}

@inproceedings{White:2009:WSDM,
  title = {Characterizing the Influence of Domain Expertise on Web Search Behavior},
  booktitle = {Proceedings of the Second ACM International Conference on Web Search and Data Mining},
  author = {White, Ryen W. and Dumais, Susan and Teevan, Jaime},
  year = 2009,
  series = {WSDM '09},
    pages={132--141}
}

@article{Mao:2018:TOIS,
  title = {How Does Domain Expertise Affect Users' Search Interaction and Outcome in Exploratory Search?},
  author = {Mao, Jiaxin and Liu, Yiqun and Kando, Noriko and Zhang, Min and Ma, Shaoping},
  year = 2018,
  journal = {ACM Transactions on Information Systems},
  volume = {36},
  number = {4},
  pages = {1--30}
}

@article{Tabatabai:2005:LISR,
  title = {How experts and novices search the Web},
  author = {Tabatabai, Diana and Shore, Bruce M.},
  year = 2005,
  journal = {Library \& Information Science Research},
  volume = {27},
  number = {2},
  pages = {222--248}
}

@inproceedings{Noh:2023:CHI,
  title = {A Study on User Perception and Experience Differences in Recommendation Results by Domain Expertise: The Case of Fashion Domains},
  booktitle = {Extended Abstracts of the 2023 CHI Conference on Human Factors in Computing Systems},
  author = {Noh, Taehyung and Yeo, Haein and Kim, Myungjin and Han, Kyungsik},
  year = 2023,
  series = {CHI EA '23},
  pages = {1--7}
}

@inproceedings{Ferrod:2021:UMAP,
  title = {Identifying users' domain expertise from dialogues},
  booktitle = {Adjunct Proceedings of the 29th ACM Conference on User Modeling, Adaptation and Personalization},
  author = {Ferrod, Roger and Cena, Federica and Di Caro, Luigi and Mana, Dario and Simeoni, Rossana Grazia},
  year = 2021,
  series = {UMAP '21},
  pages = {29--34}
}

@inproceedings{McAuley:2013:WWW,
  title = {From amateurs to connoisseurs: modeling the evolution of user expertise through online reviews},
  booktitle = {Proceedings of the 22nd international conference on World Wide Web},
  author = {McAuley, Julian John and Leskovec, Jure},
  year = 2013,
  series = {WWW '13},
  pages = {897--908}
}

@misc{Kiseleva:2015:arXiv,
  title = {The Impact of Technical Domain Expertise on Search Behavior and Task Outcome},
  author = {Kiseleva, Julia and Garc{\'i}a, Alejandro Montes and Kamps, Jaap and Spirin, Nikita},
  year = 2015,
      eprint={1512.07051},
      archivePrefix={arXiv},
      primaryClass={cs.IR}
}

@inproceedings{Yu:2018:SIGIR,
  title = {Predicting User Knowledge Gain in Informational Search Sessions},
  booktitle = {The 41st International ACM SIGIR Conference on Research \& Development in Information Retrieval},
  author = {Yu, Ran and Gadiraju, Ujwal and Holtz, Peter and Rokicki, Markus and Kemkes, Philipp and Dietze, Stefan},
  year = 2018,
  series = {SIGIR '18},
  pages = {75--84}
}

@inproceedings{Zhang:2011:SIGIR,
  title = {Predicting users' domain knowledge from search behaviors},
  booktitle = {Proceedings of the 34th international ACM SIGIR conference on Research and development in Information Retrieval},
  author = {Zhang, Xiangmin and Cole, Michael and Belkin, Nicholas},
  year = 2011,
  series = {SIGIR '11},
  pages = {1225--1226}
}

@article{Zhang:2015:JASIST,
  title = {Predicting users' domain knowledge in information retrieval using multiple regression analysis of search behaviors},
  author = {Zhang, Xiangmin and Liu, Jingjing and Cole, Michael and Belkin, Nicholas},
  year = 2015,
  journal = {Journal of the Association for Information Science and Technology},
  volume = {66},
  number = {5},
  pages = {980--1000}
}

@inproceedings{Knijnenburg:2009:RecSys,
  title = {Understanding the effect of adaptive preference elicitation methods on user satisfaction of a recommender system},
  booktitle = {Proceedings of the third ACM conference on Recommender systems},
  author = {Knijnenburg, Bart P. and Willemsen, Martijn C.},
  year = 2009,
  series = {RecSys '09},
  pages = {381--384}
}

@article{VonAhn:2006:Computer,
  title={Games with a purpose},
  author={Von Ahn, Luis},
  journal={Computer},
  volume={39},
  number={6},
  pages={92--94},
  year={2006}
}

@misc{hou2024bridging,
      title={Bridging Language and Items for Retrieval and Recommendation: Benchmarking LLMs as Semantic Encoders}, 
      author={Yupeng Hou and Jiacheng Li and Xiangjun Fu and Zhankui He and An Yan and Xiusi Chen and Julian McAuley},
      year={2026},
      eprint={2403.03952},
      archivePrefix={arXiv},
      primaryClass={cs.IR}
}

@article{Lee:2006:ASIST,
author = {Lee, Jin Ha and Renear, Allen and Smith, Linda C.},
title = {Known-Item Search: Variations on a Concept},
journal = {Proceedings of the American Society for Information Science and Technology},
volume = {43},
number = {1},
pages = {1--17},
year = {2006}
}

\end{document}